# Spatially Resolved Dynamics of Localized Spin-Wave Modes in Ferromagnetic Wires

J. P. Park, P. Eames, D. M. Engebretson, J. Berezovsky, and P. A. Crowell

*School of Physics and Astronomy, University of Minnesota, 116 Church St. SE
Minneapolis, MN 55455*

Abstract

We have observed localized spin-wave modes in individual thin-film ferromagnetic wires using time-resolved Kerr microscopy as a micron-scale spectroscopic probe. The localization is due to the partial demagnetization of a wire when an external field is applied in the plane of the film and perpendicular to the long axis of the wire. Spatially-resolved spectra demonstrate the existence of distinct modes at the edges of a rectangular wire. Spectral images clearly show the crossover of the two edge modes into a single mode in low applied fields, in agreement with the results of micromagnetic simulations.

An important aspect of the physics of nanostructures is how spin dynamics are influenced by both geometric confinement and magnetic microstructure. In addition to determining parameters of technological relevance, such as switching times in recording media,[1] the collective modes of patterned thin films also pose a problem of fundamental physical interest. They cannot be described completely as either purely magnetostatic modes[2] or as simple spin-waves.[3] In patterned thin films, the competition between the magnetostatic and exchange energies can lead to unique excitations, for which spectroscopic evidence has been found in recent Brillouin light scattering measurements on arrays of ferromagnetic wires.[4] An investigation of these modes, which can be qualitatively described as localized spin-waves, is an essential step in achieving a complete understanding of the micromagnetic dynamics of low-dimensional structures.

This Letter reports the direct observation of localized spin-wave modes in individual thin-film ferromagnetic wires using time-resolved scanning Kerr microscopy,[5-7] which we demonstrate can be used as a local spectroscopic probe. The effective confinement of these modes is produced by the demagnetizing field when the applied field is in the plane of the film and perpendicular to the axis of the wire.[4] In sufficiently large magnetic fields, these modes exist along each edge of the wire in addition to the conventional ferromagnetic resonance (FMR) mode at the center. Our measurements show a crossover from this multi-mode structure to a single mode as the applied field decreases and the barrier between the two edge modes disappears.

Permalloy ($Ni_{0.81}Fe_{0.19}$) films of thickness 18 nm were sputtered onto GaAs substrates and patterned into wires of width 2 μm and 5 μm by electron beam lithography and lift-off. A bilayer resist process was used to fabricate a lithographically defined mask with reentrant side walls. This approach produces wires with sharply defined edges. The wires are 1 mm long, and the center-to-center spacing is 15 μm. The GaAs substrates are polished to a thickness less than 30 μm and are then positioned over the center of a tapered microstrip line, with the external dc field perpendicular to the microwave magnetic field. The samples can be rotated on the stripline so that the long axes of the wires are either parallel or perpendicular to the external field, which is always in the plane of the film. We will use the terminology applied to Brillouin light scattering experiments,[8] with the understanding that the relevant wave-vectors **q** for the excitations under investigation are in the plane of the wires and perpendicular to their length. Our discussion will focus on the configuration in which $\mathbf{H_a} \parallel \mathbf{q}$, where $\mathbf{H_a}$ is the dc applied field. This is abbreviated as the BWVMS (or backward volume magnetostatic spin wave) geometry.[8] The configuration in which $\mathbf{H_a} \perp \mathbf{q}$ (i.e. the field is parallel to the wire) will be referred to as the Damon-Eshbach (DE) geometry.[9]

For the time-resolved Kerr microscopy (TRKM) measurement,[5] a 76 MHz train of 150 femtosecond pulses from a Ti:Sapphire laser ($\lambda$ = 810 nm) is split into two beams. One beam is incident on a fast photodiode, creating a current pulse of 120 – 150 ps FWHM that is discharged into the stripline. This creates an in-plane field pulse at the sample with an amplitude of 5 – 10 Oe. The second beam, used as the probe, is focused on the sample using an objective lens with a numerical aperture of 0.85, and the polar Kerr rotation of the reflected probe beam is measured with a polarization bridge. The probe spot diameter is 2 μm. The timing between the pump and probe pulses is controlled by a mechanical delay line. The signal is detected using a lock-in referenced to the pump-beam chopper, so that only the pump-induced Kerr rotation, averaged over many pulses, is recorded. Figure 1 shows the polar Kerr rotation as a function of pump-probe delay for three different positions of the probe beam on a 5 µm wide wire. These data, which are proportional to the component of the magnetization perpendicular to the plane of the film, were obtained in a field of 100 Oe applied in the BWVMS geometry. Fourier transforms of these free induction decays are shown in the right-hand panels of Fig. 1. The response at the center of the wire, shown in Fig. 1(a), is simple precession of the magnetization. However, a second, lower frequency is observed as the objective is moved towards the edge of the sample, as shown in Figs. 1(b) and 1(c).

A more complete picture is revealed by examining the full spatiotemporal images shown in Fig. 2. The panels on the left are a compilation of successive snapshots of a cross-section of a 5 µm wire, obtained in 20 or 40 ps steps. We focus first on the three lowest panels, which are images obtained in the BWVMS geometry. At 200 and 100 Oe, a clear beating pattern is observed near the edges of the wire. At 50 Oe, the beating extends over the entire sample. Spectral images are obtained by Fourier transformation of the time-domain data at each position. The spectral images, normalized by the total power at each position, are shown in the right-hand panels of Fig. 2, which clearly show distinct modes localized near the edges of the wire at 100 and 200 Oe. In contrast, images obtained in the longitudinal (DE) geometry, with the magnetic field parallel to the wire, show only a single uniform precession frequency corresponding to the lowest Damon-Eshbach mode. Time-domain and spectral images in a field of 200 Oe in the DE geometry are shown in the top panels of Fig. 2.

A detailed view of the evolution of both the center and edge modes is provided in Fig. 3, which shows a succession of spectral images at different magnetic fields for both a 2 µm and a 5 µm wire. This figure demonstrates several important facts about the localized edge modes. The confinement is strongest at the highest magnetic fields. In the case of the 2 µm wire at 400 Oe, the edge and center mode images overlap in space, although the edge mode can still be resolved spectrally as a weak satellite of the dominant center mode peak. As the magnetic field decreases, the frequencies of both the center and edge modes decrease, and the spatial extent of the edge modes grows while that of the center mode shrinks. In the case of the 2 µm wire, the modes merge below 75 Oe into a single mode spanning the entire wire. The same process appears to occur in the 5 µm wire at a lower magnetic field. At the lowest fields, the single mode observed in the 2 µm wire actually increases in frequency as the external field decreases.

The data obtained here confirm directly the interpretation of Brillouin light scattering measurements by Jorzick *et al.*,[4] who identified a non-dispersive mode in the BWVMS spectrum of wire arrays as a localized spin-wave. The images of Fig. 3 demonstrate that this is indeed the case. We now turn to a detailed interpretation of the magnetic field dependence, for which the spectral images are particularly useful. For spin-waves propagating in a thin film along the direction of the magnetic field with in-plane wave-vector **q**, the dispersion relation $v(q)$ is[10]

$$v(q) = \frac{\gamma}{2\pi} \left[ \left( H + \frac{2A}{M_S} q^2 \right) \left( H + \frac{2A}{M_S} q^2 + 4\pi M_S \left( \frac{1 - e^{-qd}}{qd} \right) \right) \right]^{1/2}, \quad (1)$$

where $\gamma$ is the gyromagnetic ratio, $H$ is the total internal field, including the demagnetizing field, $A$ is the spin-wave stiffness, $M_S$ is the saturation magnetization, and $d$ is the film thickness. We have ignored dispersion due to the out-of-plane component $q_\perp$ of the wave-vector. This is justified in our case because the first perpendicular standing spin-wave mode should occur at approximately 20 GHz, which is well outside the bandwidth of our excitation pulse. The dynamic correction to the demagnetization energy in the last term of Eq. 1 is extremely important because it leads to a minimum in the dispersion relation at non-zero $q$ in the BWVMS geometry. The uniform mode ($q = 0$) is therefore unstable with respect to the formation of smaller wavelength excitations as long as some means, such as multi-magnon scattering, is available to conserve momentum.

One approach to solving for the allowed modes of the wire follows Jorzick *et al.*, who used the WKB argument of Schlömann and Joseph.[11] This is based on the observation that the equation of motion for the magnetization vector has the same form as a Schrödinger equation, and bound states can then be found from the Bohr-Sommerfeld quantization condition

$$\int_{x_1}^{x_2(\nu)} q(\nu, H(x)) dx = n\pi. \qquad (2)$$

This is an integral equation that can be solved for $\nu$ once the turning points $x_1$ and $x_2(\nu)$ are known. Spin waves will not propagate into the demagnetized region at the edge of the wire,[4, 12] and so the turning point $x_1$ can be determined simply by calculating where the internal field drops to zero. We have done this by calculating the internal field in the simple approximation in which the poles at the edges of the wire are treated as magnetic charges. The second turning point can be determined from the position corresponding to the internal field at which there is no longer a real wave-vector associated with the frequency $\nu$. Practically, this is implemented by searching for the field that has a minimum in the dispersion relation at a given frequency $\nu$, and then using the calculated field profile to find the turning point $x_2(\nu)$. This approach allows for the solution of Eq. 2 to be automated easily.

The solutions of Eq. 2 for $n = 1$ correspond to the edge modes seen in Fig. 3. We have solved for these using the parameters $\gamma/2\pi = 2.95$ GHz/kOe, $A = 1.3 \times 10^{-6}$ erg/cm, $M_s = 700$ emu/cm$^3$, and $d = 17.5$ nm. The value for $M_S$ is the average ($700 \pm 40$ emu/cm$^3$) of FMR measurements on a witness film sample, a magnetic moment measurement by vibrating sample magnetometry (VSM), and a VSM measurement of the perpendicular saturation field. The frequencies corresponding to the $n = 1$ solutions of Eq. 2 are shown in Fig. 4(a) as the lower solid curve. The dominant contribution to the response at the center of the wire, where confinement effects are small, comes from the $q = 0$ modes of the dispersion relation. The frequencies for these uniform FMR modes, calculated from Eq. 1 using the internal field at the center of the wire, are shown for the 2 μm wire in Fig. 4(a) as the upper solid curve.

As can be seen in Fig. 4(a), there is excellent qualitative and fair quantitative agreement between the model calculations and the edge and center mode frequencies observed for the 2 μm wire above 75 Oe. However, as the applied field decreases, the experimental edge mode frequencies drop faster than the model predicts. Below 75 Oe, the effective barrier between the two edge modes disappears and the center mode frequency calculated from the model drops to zero. This occurs when the wire becomes completely demagnetized along the direction of the applied field. In order to describe the region near and below this critical field, we turn instead to a micromagnetic description of the dynamics using the discretized Landau-Lifshitz Gilbert (LLG) equation,[13]

$$(1+\alpha^2)\frac{\partial \mathbf{M}_i}{\partial t} = -\gamma(\mathbf{M}_i \times \mathbf{H}_{eff,i}) - \frac{\gamma\alpha}{M_S}\mathbf{M}_i \times (\mathbf{M}_i \times \mathbf{H}_{eff,i}), \quad (3)$$

where $\mathbf{H}_{eff}$ is the total effective field, which includes the applied field, demagnetizing field, and exchange field, $\alpha$ is the Gilbert damping parameter, and the subscript $i$ indexes square cells, inside of which the magnetization and effective field are assumed to be uniform. The solution of Eq. (3) was implemented numerically for 2 µm by 32 µm wires using the Object-Oriented Micromagnetic Framework (OOMMF)[14] and a cell size of 20 nm. The system was first relaxed from a fully magnetized state with a large damping parameter $\alpha = 0.5$. The relaxed state was then taken as the initial condition for the next stage, which was the response of the system to the magnetic field pulse. This was undertaken with a realistic damping parameter $\alpha = 0.008$. A Gaussian pulse with a FWHM of 150 ps and an amplitude of 10 Oe was applied, and the response over the next 5 ns was calculated using 10 ps steps during the pulse and 50 ps after the pulse. A spectral image of a cross-section of the wire was then constructed in a manner analogous to our analysis of the experimental data. An image obtained at 200 Oe is shown in Fig. 4(b), in which the edge and center modes are clearly visible. Frequencies of the edge and center modes determined from the LLG calculations are shown in Fig. 4(a) as open squares.

At high fields, the edge mode frequencies from the LLG calculation agree well with the WKB approach, in spite of the fact that the WKB approximation is not strictly valid for the lowest bound states. The LLG values for the center-mode frequencies are also in fair agreement with those from the simple analytical approach. Significantly, the LLG calculations show a single dominant mode at low fields, below the demagnetization field, as can be seen in the spectral image at 25 Oe shown in Fig. 4(c). This is in qualitative agreement with the experimental results shown in Fig. 3, although the experimental frequency is lower. Essentially, the low-field mode is due to precession about the shape anisotropy field, which is parallel to the wire. There is no confinement in this field regime, because the internal field perpendicular to the long axis is uniformly zero across the wire, and hence only a single mode exists. We note that the experimental edge mode frequencies are consistently lower than all sets of theoretical predictions below 150 Oe. We do not have a detailed explanation for this observation, although it is reasonable that larger deviations from the predicted internal field profiles should occur as the wire becomes demagnetized.

In summary, we have imaged localized spin wave modes in isolated ferromagnetic wires using time-resolved Kerr microscopy as a micron-scale spectroscopic probe. Analytical calculations and a micromagnetic approach provide an excellent qualitative description of the observed dynamics. The spectroscopic imaging approach adopted here can be generalized to other inhomogeneous magnetic systems, including lithographically patterned films as well as domain structures.

This work was supported by NSF DMR 99-83777, the Research Corporation, the Alfred P. Sloan Foundation, the University of Minnesota MRSEC (DMR 98-09364), and the Minnesota Supercomputing Institute. We acknowledge C. E. Campbell and M. Yan for numerous helpful discussions.

**Figure Captions**

Fig. 1: Time response (left) and frequency response (right) of the polar Kerr angle, which is proportional to the out-of-plane component of the magnetization, after a 120 ps field pulse for a 5 µm wide wire with an external field of 100 Oe applied perpendicular to its long axis. The three panels show the response measured at three different locations on the wire: a) at the center, b) 1.5 µm from the center, and c) 2.3 µm from the center, which corresponds to the edge of the wire. The applied field geometry is shown in the inset of (c).

Fig. 2: The out-of-plane component of the magnetization is shown in time-domain images of a cross-section of a 5 µm wire. The three lower rows show images in fields of 50, 100, and 200 Oe applied perpendicular to the long axis of the wire. Black and white indicate positive and negative magnetization. The right-hand panels show the same cross-section in the frequency domain, normalized the by the total spectral power at each position. The top panels, labeled DE, show images at 200 Oe in the Damon-Eshbach geometry, with the applied field along the axis of the wire. Note that the edge modes do not appear in the DE geometry.

Fig. 3: Frequency domain images of cross-sections of the 2 µm and 5 µm wires at various magnetic fields. Note the crossover to a single mode in the 2 µm wire in applied fields below 75 Oe.

Fig. 4: (a) Experimental and theoretical results for the center and edge modes. The experimental frequencies of the edge and center modes for the 2 μm wire are shown as solid triangles. Only a single mode exists below 75 Oe. The two solid curves are the theoretical values for the center and edge modes calculated from the dispersion relation as described in the text. The open squares are the mode frequencies calculated using a full Landau-Lifshitz-Gilbert (LLG) simulation. (b) The frequency domain response of a cross-section of the wire at 200 Oe calculated by the LLG method. (c) The response calculated at 25 Oe.

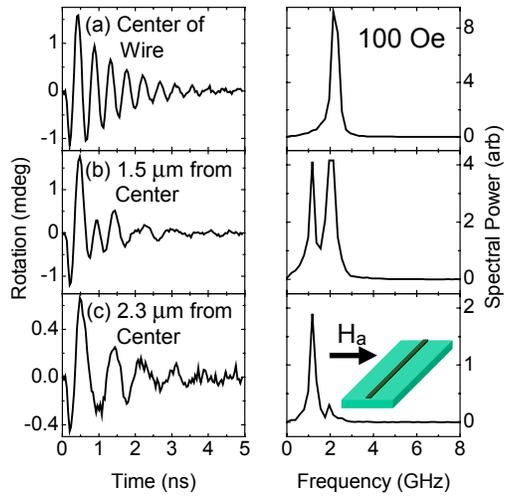

Figure 1
Park *et al.*

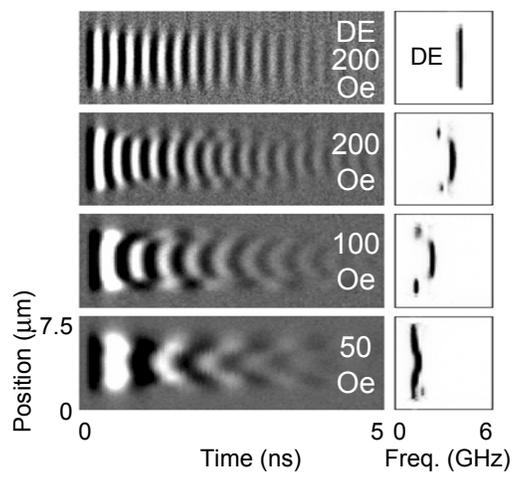

Figure 2
Park *et al.*

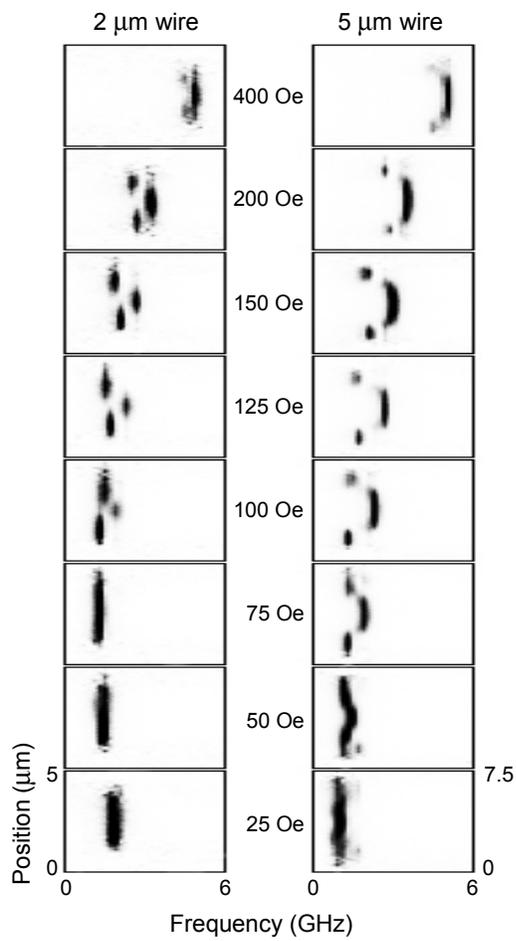

Figure 3
Park *et al*.

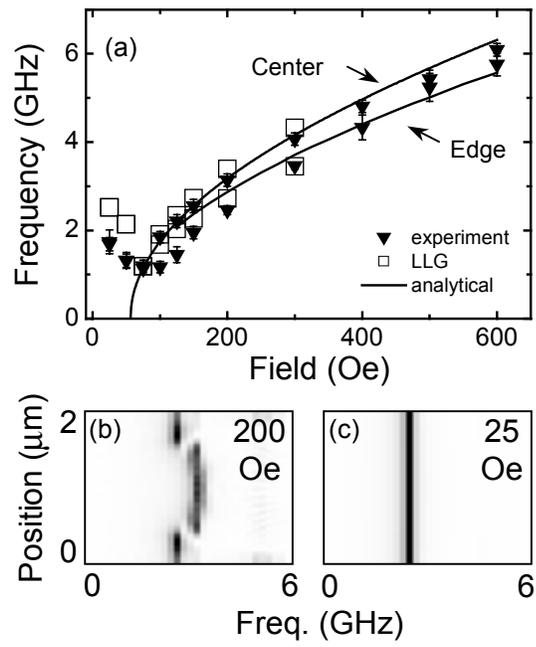

Figure 4
Park et al.